\documentclass[american,english]{IEEEtran}
\usepackage[T1]{fontenc}
\usepackage[latin9]{inputenc}
\usepackage{color}
\usepackage{units}
\usepackage{mathtools}
\usepackage{amsmath}
\usepackage{amssymb}
\usepackage{cancel}
\usepackage{mathdots}
\usepackage{stmaryrd}
\usepackage{stackrel}
\usepackage{graphicx}
\PassOptionsToPackage{version=3}{mhchem}
\usepackage{mhchem}
\usepackage{esint}

\makeatletter
\usepackage{colortbl}
\usepackage{cite}

\usepackage[labelsep=endash]{caption}
\usepackage{flushend}
\usepackage{stfloats}
\usepackage[margin=8pt,font=footnotesize,labelfont=md,labelsep=colon]{caption}

\usepackage{multicol}
\usepackage{float}
\usepackage{lipsum}
\usepackage[figurename=Fig.]{caption}

\makeatother

\usepackage{babel}
\begin{document}
\title{Reconfigurable Intelligent Surface Enabled IoT Networks in Generalized
Fading Channels}
\author{\textcolor{black}{\normalsize{}Abubakar U. Makarfi}\textit{\textcolor{black}{\normalsize{}$^{1}$,}}\textcolor{black}{\normalsize{}
Khaled M. Rabie}\textit{\textcolor{black}{\normalsize{}$^{2}$}}\textcolor{black}{\normalsize{},
Omprakash Kaiwartya}\textit{\textcolor{black}{\normalsize{}$^{3}$}}\textcolor{black}{\normalsize{},
Osamah S. Badarneh}\textit{\textcolor{black}{\normalsize{}$^{4}$,
}}\textcolor{black}{\normalsize{}Xingwang Li}\textit{\textcolor{black}{\normalsize{}$^{5}$,}}\textcolor{black}{\normalsize{}
Rupak Kharel}\textit{\textcolor{black}{\normalsize{}$^{1}$}}\textcolor{black}{\normalsize{}}\\
\textcolor{black}{\normalsize{}$^{1}$Department of Computing and
Mathematics, Manchester Metropolitan University, UK}\\
\textit{\textcolor{black}{\normalsize{}$^{2}$}}\textcolor{black}{\normalsize{}Department
of Engineering, Manchester Metropolitan University, UK}\\
\textit{\textcolor{black}{\normalsize{}$^{3}$}}\textcolor{black}{\normalsize{}School
of Science and Technology, Nottingham Trent University, UK }\\
\textit{\textcolor{black}{\normalsize{}$^{4}$}}\textcolor{black}{\normalsize{}Electrical
and Communication Engineering Department, German-Jordanian University,
Amman, Jordan}\\
\textit{\textcolor{black}{\normalsize{}$^{5}$}}\textcolor{black}{\normalsize{}School
of Physics and Electronic Information Engineering, Henan Polytechnic
University, China }\\
\textcolor{black}{\normalsize{}Emails:\{a.makarfi, r.kharel, k.rabie\}@mmu.ac.uk;
omprakash.kaiwartya@ntu.ac.uk; }\\
\textcolor{black}{\normalsize{}Osamah.Badarneh@gju.edu.jo; lixingwang@hpu.edu.cn. }}

\maketitle
\selectlanguage{american}%
\textcolor{black}{\thispagestyle{empty}}
\selectlanguage{english}%
\begin{abstract}
This paper studies an Internet-of-Things (IoT) network employing a
reconfigurable intelligent surface (RIS) over generalized fading channels.
Inspired by the promising potential of RIS-based transmission, we
investigate a RIS-enabled IoT network with the source node employing
a RIS-based access point. The system is modelled with reference to
a receiver-transmitter pair and the Fisher-Snedecor $\mathcal{F}$
model is adopted to analyse the composite fading and shadowing channel.\textcolor{black}{{}
}Closed-form expressions are derived for the system with regards to
the average capacity, average bit error rate (BER) and outage probability.
Monte-Carlo simulations are provided throughout to validate the results.
The results investigated and reported in this study extend early results
reported in the emerging literature on RIS-enabled technologies and
provides a framework for the evaluation of a basic RIS-enabled IoT
network over the most common multipath fading channels. The results
indicate the clear benefit of employing a RIS-enabled access point,
as well as the versatility of the derived expressions in analysing
the effects of fading and shadowing on the network. The results further
demonstrate that for a RIS-enabled IoT network, there is the need
to balance between the cost and benefit of increasing the RIS cells
against other parameters such as increasing transmit power, especially
at low SNR and/or high to moderate fading/shadowing severity.
\end{abstract}


\section{\textcolor{black}{Introduction}}

\textcolor{black}{Reconfigurable intelligent surfaces (RIS) are one
of the emerging technologies, aimed at enabling the concept of ``smart
radio environments'' in beyond 5G communication networks. The idea
behind such technologies is to control }the propagation environment
in order to improve signal quality and coverage \textcolor{black}{\cite{Renzo2019,Liaskos_metasurface}.
RISs are man-made surfaces of electromagnetic material that are electronically
controlled with integrated electronics and have unique wireless communication
capabilities \cite{Basar2019WirelessCT}.} RIS-based transmission
schemes have several benefits and key features. Ideally, they are
conceptualised to be nearly passive with dedicated energy sources,
the surface can shape the wave incident upon it, they have full-band
response and easily deployable on different surfaces like buildings,
vehicles or indoor spaces. Additionally, they are not affected by
receiver noise and have no power amplifiers, thus, they do not amplify
nor introduce noise when reflecting signals \textcolor{black}{\cite{Basar2019WirelessCT,Basar_xmsn_LIS}}. 

RIS-enabled applications have recently been investigated with respect
to signal-to-noise ratio (SNR) maximisation \cite{Basar_xmsn_LIS},
improving signal coverage \cite{subrt}, improving massive MIMO systems
\cite{MIMO_LIS}, beamforming optimisation \cite{Wu_beam_opt,Wu_beamform,7510962},
as well as multi-user networks\cite{LIS_multi_user}. On the other
hand, important 5G and beyond technologies such as Internet-of-Things
(IoT) devices are yet to be explored with respect to RIS-enabled schemes,
even though the IoT paradigm has recently received much attention
in the literature \cite{FD_ambroziak,measuresVCPS,fuzzyVCPS,EH_D2D,iot_d2d_lte_elmesalawy}.
IoT devices take advantage of short communication ranges to reduce
latency, while optimising spectrum efficiency and energy efficiency
\cite{d2d_iot_anajemba,iot_d2d_lte_elmesalawy,IoT_book}, which are
parameters that can be further enhances with RIS-enabled technologies
\cite{Basar2019WirelessCT}. 

Regarding the channel characteristics, generalised fading distributions
have been proposed that include or closely approximate the most common
fading distributions as special cases \cite{Yacoub_Alp_mu,Yacoub_eta_mu,Yacoub_kappa_mu,Li_k_u},
thus providing ease and flexibility of analysis. For example, a wireless
network over generalised $\alpha-\mu$ fading channels was studied
for the particular cases of the Weibull, Nakagami$-m$ and Rayleigh
channels in \cite{GalymAccess18}, while a network over generalized
$\kappa-\mu$ fading for the special cases of the Rician, Nakagami-$m$
and Rayleigh channels was studied in \cite{EH_k_u_letter}.\textcolor{blue}{{}
}\textcolor{black}{More recently, it was shown in \cite{FisherModel17}
that the Fisher-Snedecor $\mathcal{F}$ composite fading model outperforms
most other generalized fading models because it provides accurate
modeling and characterisation of the simultaneous occurrence of multipath
fading and shadowing. Furthermore, the Fisher-Snedecor $\mathcal{F}$
model includes several fading distributions as special cases, such
as Nakagami-$m$ ($m_{s}\rightarrow\infty$), Rayleigh ($m_{s}\rightarrow\infty,$
$m=1$) and one-sided Gaussian distribution ($m_{s}\rightarrow\infty,$
$m=\nicefrac{1}{2}$), while being more mathematically tractable.
Thus, a key motivation for this study.}

\textcolor{black}{In light of the foregoing, this study seeks to contribute
the following. The analysis of a RIS-based IoT network over composite
fading and shadowing Fisher-Snedecor channels, where a reference source
node employs a RIS configured access point for transmission. This
allows for the derivation of expressions for three metrics of performance
analysis, namely; the average capacity, the bit error rate (BER) and
the outage probability of the system. To the best of our knowledge,
this is the first analysis of a RIS-based IoT network over composite
fading and shadowing. The expressions are derived in closed-form and
Monte Carlo simulations are provided throughout to verify the accuracy
of our analysis. }Additionally, asymptotic expressions of the system
metrics are derived and compared with the exact solutions. These expressions
were shown to provide further insight to the network analysis, while
more tractable analytically. The results indicate the clear benefit
of employing a RIS-enabled access point, with regards to the average
capacity, BER and outage probability, as well as the versatility of
the derived expressions in analysing the effects of fading and shadowing
on the network. The results further demonstrate that for a RIS-enabled
IoT network, there is the need to balance between the cost and benefit
of increasing the RIS cells against other parameters such as increasing
transmit power, especially at low SNR and/or high to moderate fading/shadowing
severity. 

The paper is organised as follows. In Section \ref{sec:sys-model},
we describe the system model under study. Thereafter, in Section \ref{sec:perf-anal},
we derive expressions for efficient computation of the average capacity,
BER and outage probability of the system. Finally, in Sections \ref{sec:Results}
and \ref{sec:Conclusions}, we present the results and outline the
main conclusions, respectively. 

\section{System Model\label{sec:sys-model}}

\begin{figure}[th]
\begin{centering}
\includegraphics[scale=0.45]{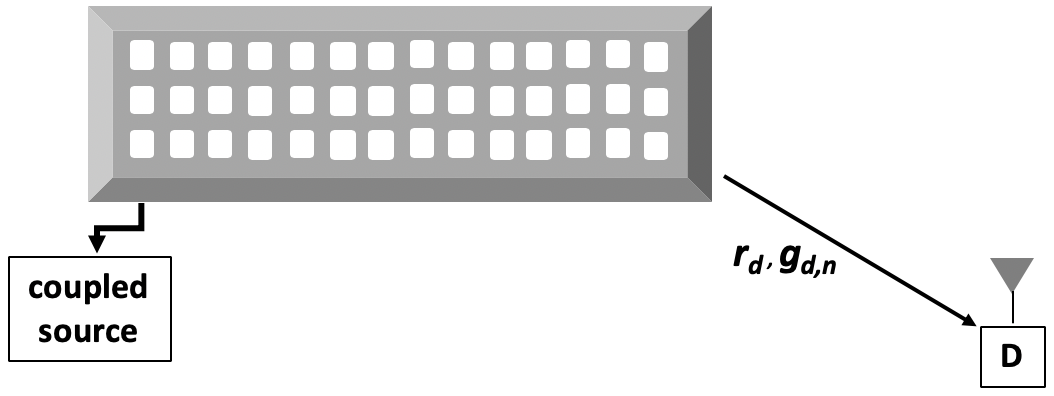}
\par\end{centering}
\caption{Source node RIS-based configuration employed as an access point.\label{fig:sys-mod}}
\end{figure}

We consider a network of IoT nodes, as illustrated in Fig. \ref{fig:sys-mod}.
Each transmitter-receiver pair consists of a simple source node ($S$)
and a destination node ($D$) for the transmitted information. The
node $S$ is assumed to employ a RIS (consisting of passive reflector
elements) in the form of an AP to communicate over the network\footnote{Initial proposal and results for such a configuration of intelligent
surfaces were reported in \cite{Basar_xmsn_LIS}.}. As shown in the block diagram, the RIS can be connected over a wired
link or optical fiber for direct transmission from $S$, and can support
transmission without RF processing. For the system considered, we
assume an intelligent AP with the RIS having knowledge of channel
phase terms, such that the RIS-induced phases can be adjusted to maximise
the received SNR through appropriate phase cancellations and proper
alignment of reflected signals from the intelligent surface.

The received signal at $D$ can be represented as
\begin{align}
y_{D} & =\left[\sum_{n=1}^{N}h_{D,n}e^{-j\phi_{n}}\right]x+w_{D},\label{eq:yD}
\end{align}
where $x$ represents the transmitted signal by $S$ with power $P_{s}$
and $w_{D}$ the additive white Gaussian noise (AWGN) at $D$. Without
loss of generality, we denote the power spectral density of the AWGN
as $N_{0}$. The term $\phi_{n}$ in (\ref{eq:yD}) is the reconfigurable
phase induced by the $n$th reflector of the RIS, which through phase
matching, the SNR of the received signals can be maximised\footnote{For the sake of brevity, the reader is referred to \cite{Basar_xmsn_LIS},
for details of phase cancellation techniques.}. The term $h_{D,n}=\sqrt{g_{d,n}r_{d}^{-\beta}}$, is the channel
coefficient from $S$-to-$D$ with distance $r_{d}$, path-loss exponent
$\beta$ and channel gain $g_{d,n}$ modelled as independently distributed
Fisher-Snedecor $\mathcal{F}$ RVs with the following PDF and CDF,
respectively \cite{FisherModel17} 
\textcolor{black}{
\begin{align}
f\left(g_{n}\right) & =\Upsilon_{n}g_{n}^{-1}\textrm{G}_{1,1}^{1,1}\biggl[\Lambda_{n}\,g_{n}\Biggl|\negthickspace\begin{array}{c}
1-m_{s_{n}}\\
m_{n}
\end{array}\negthickspace\biggr],m_{s_{n}}>1,\label{eq:pdfX}
\end{align}
}and 
\begin{equation}
F_{g_{n}}\left(v\right)=\Upsilon_{n}\,\textrm{G}_{2,2}^{1,2}\biggl[\Lambda_{n}v\Biggl|\negthickspace\begin{array}{c}
1-m_{s_{n}},\,1\\
m_{n},\,0
\end{array}\negthickspace\biggr],\label{eq:CDFX}
\end{equation}

\noindent \textcolor{black}{where $n\in\left\{ 1,2,\ldots,N\right\} $,
$\Upsilon_{n}=\frac{1}{\Gamma\left(m_{n}\right)\Gamma\left(m_{s_{n}}\right)}$
and $\Lambda_{n}=\frac{m_{n}}{\left(m_{s_{n}}-1\right)\bar{g}_{n}}$.
The terms $\textrm{G}_{u,v}^{s,t}\left[\cdotp\right]$ is the Meijer\textquoteright s
G-function \cite[Eq. (8.3.1.1)]{bookV3}, $\Gamma\left(z\right)=\int_{0}^{\infty}t^{z-1}e^{-t}dt$
}is the gamma function \cite[Eq. (8.310)]{book2}\textcolor{black}{and
$\bar{g}_{n}=\mathbb{E}\left[g_{n}\right]$ is the mean power of the
RV $g$ with $\mathbb{E}\left[\cdotp\right]$ denoting expectation.
$m_{n}$ and $m_{s_{n}}$ represent the fading severity and shadowing
parameters of the $n$-th RV, respectively.}

Based on (\ref{eq:yD}), the instanstaneous SNR at $D$ is given by
\begin{equation}
\gamma_{D}=\frac{\sum_{n=1}^{N}P_{s}\mid h_{D,n}\mid^{2}}{N_{0}}.\label{eq:sinr-d}
\end{equation}

\section{Mathematical Analysis\label{sec:perf-anal}}

In this section, we derive analytical expressions for various performance
metrics of the system under consideration.

\subsection{Average Capacity Analysis \label{sec:cap-anal}}

In this section, we derive analytical expressions for the average
capacity over Fisher-Snedecor $\mathcal{F}$ composite fading. The
capacity is defined by 
\begin{equation}
C_{D}=\log_{2}\left(1+\gamma_{D}\right),\label{eq:cs-defined}
\end{equation}
where $\gamma_{D}$ is the SNR at $D$. 

The average capacity is given by 
\begin{align}
\overline{C}_{D} & =\mathbb{E}\left[\log_{2}\left(1+\gamma_{D}\right)\right]\nonumber \\
 & =\stackrel[0]{\infty}{\int}\log_{2}\left(1+\frac{P_{s}r_{d}^{-\beta}}{N_{0}}g_{D}\right)f\left(g_{D}\right)\textrm{d}g_{D},\label{eq:av-cap-1}
\end{align}
where $g_{D}=\sum_{n=1}^{N}g_{d,n}$. As can be observed, the RV $g_{D}$
is a sum of independent Fisher-Snedecor $\mathcal{F}$ distributed
RVs, therefore, from (\ref{eq:pdfX}) and \cite[Eq. (7)]{badarneh2019sum},
the PDF is given by 
\begin{multline}
f\left(g\right)=\frac{1}{gB\left(Nm,Nm_{s}\right)}\left(\frac{gm}{Nm_{s}}\right)^{Nm}\\
\times{}_{\phantom{}2}F_{1}\left(N\left(m+m_{s}\right),Nm;Nm;-\frac{gm}{Nm_{s}}\right),\label{eq:sum-pdf-1}
\end{multline}
where $_{\phantom{}2}F_{1}\left(\alpha;\beta;\gamma;z\right)$ is
the Gauss hypergeometric function \cite[Eq. (9.111)]{book2} and $B\left(.,.\right)$
is the beta function \cite[Eq. (8.384.1)]{book2}. To proceed, we
express the PDF (\ref{eq:sum-pdf-1}) in a more tractable form, by
re-writing the PDF in an alternate form. 

Using the relation $B\left(x,y\right)=\frac{\Gamma\left(x\right)\Gamma\left(y\right)}{\Gamma\left(x+y\right)}$
\cite[Eq. (8.384.1)]{book2} and $_{\phantom{}2}F_{1}\left(\alpha;\beta;\gamma;-z\right)=\frac{\Gamma\left(\gamma\right)z}{\Gamma\left(\alpha\right)\Gamma\left(\beta\right)}\textrm{G}_{2,2}^{1,2}\left(z\Biggl|\negthickspace\begin{array}{c}
-\alpha,\,-\beta\\
-1,\,-\gamma
\end{array}\negthickspace\right)$ \cite[Eq. (8.384.1)]{book2} with some algebraic manipulations, the
PDF can be rewritten as

\begin{multline}
f\left(g\right)=\frac{g^{Nm}}{\Gamma\left(Nm\right)\Gamma\left(Nm_{s}\right)}\left(\frac{m}{Nm_{s}}\right)^{Nm+1}\\
\times\textrm{G}_{2,2}^{1,2}\left(\frac{gm}{Nm_{s}}\Biggl|\negthickspace\begin{array}{c}
-N\left(m+m_{s}\right),\,-Nm\\
-1,\,-Nm
\end{array}\negthickspace\right).\label{eq:sum-pdf-2}
\end{multline}

With the appropriate change of variables and representing the logarithmic
function in terms of the Meijer G-function with the aid of \cite[Eq. (11)]{Adamchik90},
i.e., 
\begin{equation}
\textrm{ln}\left(1+z\right)=G_{2,2}^{1,2}\Bigl[z\Bigl|\!\begin{array}{c}
1,1\\
1,0
\end{array}\!\Bigr],\label{eq:meijerG-log}
\end{equation}
we can re-express (\ref{eq:av-cap-1}) as

\begin{align}
\overline{C}_{D} & =\frac{\Lambda}{\textrm{ln}\left(2\right)}\stackrel[0]{\infty}{\int}g_{D}^{Nm}\,\textrm{G}_{2,2}^{1,2}\left(\eta g_{D}\biggl|\negthickspace\begin{array}{c}
1,1\\
1,0
\end{array}\right)\nonumber \\
 & \qquad\qquad\times\textrm{G}_{2,2}^{1,2}\left(\xi g_{D}\Biggl|\negthickspace\begin{array}{c}
-N\left(m+m_{s}\right),\,-Nm\\
-1,\,-Nm
\end{array}\negthickspace\right)\textrm{d}g_{D}.\nonumber \\
 & \stackrel{\left(a\right)}{=}\frac{\Lambda}{\textrm{ln}\left(2\right)}\stackrel[0]{\infty}{\int}\textrm{G}_{2,2}^{1,2}\left(\eta g_{D}\biggl|\negthickspace\begin{array}{c}
1,1\\
1,0
\end{array}\right)\nonumber \\
 & \qquad\qquad\qquad\quad\times\textrm{G}_{2,2}^{1,2}\left(\xi g_{D}\Biggl|\negthickspace\begin{array}{c}
-Nm_{s},\,0\\
Nm-1,\,0
\end{array}\negthickspace\right)\textrm{d}g_{D},\label{eq:av-cap-2}
\end{align}
where $\left(a\right)$ was obtained using \cite[Eq. (9.31.5)]{book2}.
The terms $\xi=\frac{m}{Nm_{s}}$ , $\eta=\frac{P_{s}r_{d}^{-\beta}}{N_{0}}$
and $\Lambda=\frac{\xi}{\Gamma\left(Nm\right)\Gamma\left(Nm_{s}\right)}.$ 

The integral in (\ref{eq:av-cap-2}) can be evaluated using \cite[Eq. (7.811.1)]{book2},
to obtain the average capacity as

\begin{equation}
\overline{C}_{D}=\frac{\Lambda}{\xi\textrm{ln}\left(2\right)}\,\textrm{G}_{4,4}^{3,3}\biggl[\,\frac{\eta}{\xi}\biggl|\negthickspace\begin{array}{c}
1-Nm,0,1,1\\
Nm_{s},0,1,0
\end{array}\negthickspace\biggr].\label{eq:av-cap-3}
\end{equation}

\subsection{Bit Error Rate Analysis\label{subsec:BER-anal}}

In this section, we analyse the BER over Fisher-Snedecor $\mathcal{F}$
composite fading. The average BER is defined as
\begin{equation}
\overline{P}_{e}=\stackrel[0]{\infty}{\int}Q\left(\sqrt{2\eta\lambda g_{D}}\right)f\left(g_{D}\right)\textrm{d}g_{D},\label{eq:av-ber-1}
\end{equation}
where \textbf{$Q\left(x\right)=\frac{1}{\sqrt{2\pi}}\stackrel[x]{\infty}{\int}\exp\left(-\frac{u^{2}}{2}\right)$}d$u$
is the Gaussian $Q$-function. The term $\lambda$ is a modulation
constant, with $\lambda=\left\{ 0.5,1\right\} $ for binary frequency
shift keying (BFSK) and binary phase shift keying (BPSK), respectively. 

Using the relation $Q\left(x\right)=\frac{1}{2}\textrm{erfc}\left(\frac{x}{\sqrt{2}}\right)$,
where $\textrm{erfc}\left(x\right)$ is the complementary error function
and expressing the error function in terms of the Meijer G-function,
i.e., 
\begin{equation}
\textrm{erfc}\left(\sqrt{z}\right)=\frac{1}{\sqrt{\pi}}G_{1,2}^{2,0}\Bigl[z\Bigl|\!\begin{array}{c}
1\\
0,0.5
\end{array}\!\Bigr],\label{eq:meijerG-erfc}
\end{equation}
we can re-express (\ref{eq:av-ber-1}) as

\begin{align}
\overline{P}_{e} & =\stackrel[0]{\infty}{\int}\frac{1}{2}\textrm{erfc}\left(\sqrt{\eta\lambda g_{D}}\right)f\left(g_{D}\right)\textrm{d}g_{D}\nonumber \\
 & \stackrel{\left(b\right)}{=}\stackrel[0]{\infty}{\int}\frac{\Lambda}{2\sqrt{\pi}}g_{D}^{Nm}G_{1,2}^{2,0}\left(\eta\lambda g_{D}\Bigl|\!\begin{array}{c}
1\\
0,0.5
\end{array}\!\right)\nonumber \\
 & \qquad\times\textrm{G}_{2,2}^{1,2}\left(\xi g_{D}\Biggl|\negthickspace\begin{array}{c}
-N\left(m+m_{s}\right),\,-Nm\\
-1,\,-Nm
\end{array}\negthickspace\right)\textrm{d}g_{D}.\nonumber \\
 & \stackrel{\left(c\right)}{=}\stackrel[0]{\infty}{\int}\frac{\Lambda}{2\sqrt{\pi}}G_{1,2}^{2,0}\left(\eta\lambda g_{D}\Bigl|\!\begin{array}{c}
1\\
0,0.5
\end{array}\!\right)\nonumber \\
 & \qquad\times\textrm{G}_{2,2}^{1,2}\left(\xi g_{D}\Biggl|\negthickspace\begin{array}{c}
-Nm_{s},\,0\\
Nm-1,\,0
\end{array}\negthickspace\right)\textrm{d}g_{D},\label{eq:av-ber-2}
\end{align}
where $\left(b\right)$ was obtained using (\ref{eq:sum-pdf-2}) and
(\ref{eq:meijerG-erfc}), while $\left(c\right)$ was obtained using
\cite[Eq. (9.31.5)]{book2}. 

Using \cite[Eq. (7.811.1)]{book2}, the average BER can be obtained
by evaluating the integral in (\ref{eq:av-ber-2}) as

\begin{equation}
\overline{P}_{e}=\frac{\Lambda}{\eta\lambda2\sqrt{\pi}}\,\textrm{G}_{4,3}^{1,4}\biggl[\,\frac{\xi}{\eta\lambda}\biggl|\negthickspace\begin{array}{c}
0,-\frac{1}{2},-Nm_{s},0\\
Nm-1,0,-1
\end{array}\negthickspace\biggr].\label{eq:av-cap-3-1}
\end{equation}

\subsection{Outage Probability Analysis\label{subsec:OP-Anal} }

In this section, we derive analytical expressions for the outage probability
over Fisher-Snedecor $\mathcal{F}$ composite fading. The outage probability
$P_{\textrm{out}}$, is defined as the probability that the received
SNR falls below a certain threshold value denoted here as $\gamma_{\textrm{th}}$.
For the considered network, the outage probability is given by
\begin{align}
P_{\textrm{out}} & =\Pr\left\{ \gamma_{D}<\gamma_{\textrm{th}}\right\} ,\nonumber \\
 & =\stackrel[0]{\gamma_{\textrm{th}}}{\int}f\left(\gamma_{D}\right)\textrm{d}\gamma_{D}\label{eq:pout-1}
\end{align}
where $\gamma_{D}$ is defined in (\ref{eq:sinr-d}). It can be observed
that $\gamma_{D}$ follows a distribution of the sum of independent
Fisher-Snedecor $\mathcal{F}$ distributed RVs, with the PDF $f\left(g_{D}\right)$
defined in (\ref{eq:sum-pdf-2}). Then from (\ref{eq:pout-1}), $P_{\textrm{out}}$
is the CDF of $\gamma_{D}$, which can be evaluated using \cite[Eq. (9)]{badarneh2019sum}
as
\begin{align}
P_{\textrm{out}} & =\frac{\Gamma\left(Nm+Nm_{s}\right)}{\Gamma\left(1+Nm\right)\Gamma\left(Nm_{s}\right)}\left(\frac{\gamma_{\textrm{th}}m}{\eta Nm_{s}}\right)^{Nm}\nonumber \\
 & \times{}_{\phantom{}2}F_{1}\left(N\left(m+m_{s}\right),Nm;1+Nm;-\frac{\gamma_{\textrm{th}}m}{\eta Nm_{s}}\right),\label{eq:pout-2}
\end{align}
 where $\eta=\frac{P_{s}r_{d}^{-\beta}}{N_{0}}$. Note that (\ref{eq:pout-2})
converges if $\mid\frac{\gamma_{\textrm{th}}m}{\eta Nm_{s}}\mid\:<1$. 

\subsection{Asymptotic Analysis\label{subsec:asympt-anal}}

The asymptotic analysis is anchored in obtaining expressions in higher
SNR regimes to gain further insight in the system performance. In
this subsection, asymptotic expressions for the average capacity,
average BER and outage probability will be derived.

\subsubsection{Average Capacity}

The asymptotic capacity can be obtained to gain further insight at
higher SNR regimes. When $g_{D}$ is sufficiently large, then $\ln\left(1+z\right)\simeq\ln\left(z\right).$
Next, from \cite[Eq. (8)]{badarneh2019sum}, we employ an alternative
form of (\ref{eq:sum-pdf-1}) for the PDF of $g_{D}$ given by 
\begin{equation}
f\left(g\right)=\frac{\left(\frac{gm}{Nm_{s}}\right)^{Nm}}{gB\left(Nm,Nm_{s}\right)}\left(1+\frac{gm}{Nm_{s}}\right)^{-N\left(m+m_{s}\right)}.\label{eq:sum-pdf-3}
\end{equation}

Using the definition of the average capacity (\ref{eq:av-cap-1})
and the PDF (\ref{eq:sum-pdf-3}), the asymptotic capacity is given
by
\begin{align}
\overline{C}_{D}^{asy} & =\stackrel[0]{\infty}{\int}\log_{2}\left(\eta g_{D}\right)f\left(g_{D}\right)\textrm{d}g_{D},\nonumber \\
 & =\frac{\left(\frac{m}{Nm_{s}}\right)^{Nm}}{\ln\left(2\right)B\left(Nm,Nm_{s}\right)}\stackrel[0]{\infty}{\int}\frac{g_{D}^{Nm-1}\ln\left(\eta g_{D}\right)}{\left(1+\frac{g_{D}m}{Nm_{s}}\right)^{N\left(m+m_{s}\right)}}\textrm{d}g_{D}.\label{eq:c-av-asympt-1}
\end{align}

By using \cite[Eq. (2.6.4.7)]{prudnikovV1}, the integral in (\ref{eq:c-av-asympt-1})
can be evaluated to obtain the asymptotic average capacity as
\begin{equation}
\overline{C}_{D}^{asy}\backsimeq\frac{\ln\left(\frac{\eta}{\xi}\right)+\psi\left(Nm\right)-\psi\left(Nm_{s}\right)}{\ln\left(2\right)},\label{eq:c-av-asympt-2}
\end{equation}
where $\psi\left(z\right)=\frac{\textrm{d log}\left(\Gamma\left(z\right)\right)}{\textrm{d}z}$
is the psi function \cite[Eq. (8.360.1)]{book2}. 

\subsubsection{Average BER}

From (\ref{eq:sum-pdf-1}), the PDF of $g_{D}$ at the origin can
be expressed as
\begin{equation}
f\left(g\right)=\frac{g^{Nm-1}}{B\left(Nm,Nm_{s}\right)}\left(\frac{m}{Nm_{s}}\right)^{Nm},\label{eq:pdf-sum-4}
\end{equation}
where we have used the hypergeometric function identity $_{\phantom{}2}F_{1}\left(\alpha;\beta;\gamma;0\right)=1.$
Thus, from the definition of the average BER (\ref{eq:av-ber-1}),
(\ref{eq:meijerG-erfc}) and the PDF (\ref{eq:pdf-sum-4}), we can
express the asymptotic BER as
\begin{align}
\overline{P}_{e}^{asy} & \backsimeq\frac{1}{2\sqrt{\pi}B\left(Nm,Nm_{s}\right)}\left(\frac{m}{Nm_{s}}\right)^{Nm}\nonumber \\
 & \qquad\times\stackrel[0]{\infty}{\int}g_{D}^{Nm-1}G_{1,2}^{2,0}\left(\eta\lambda g_{D}\Bigl|\!\begin{array}{c}
1\\
0,0.5
\end{array}\!\right)\textrm{d}g_{D}.\label{eq:av-ber-asym-1}
\end{align}

Finally, the asymptotic BER can be obtained by evaluating the integral
in (\ref{eq:av-ber-asym-1}), with the aid of \cite[Eq. (7.811.4)]{book2},
to obtain the expression as
\begin{equation}
\overline{P}_{e}^{asy}\backsimeq\frac{\Gamma\left(\frac{1}{2}+Nm\right)}{2\sqrt{\pi}B\left(Nm,Nm_{s}\right)Nm}\left(\frac{m}{\eta\lambda Nm_{s}}\right)^{Nm}.\label{eq:av-ber-asym-2}
\end{equation}

\subsubsection{Outage Probability}

The asymptotic outage probability can be obtained when $\frac{P_{s}}{N_{0}}\rightarrow\infty$.
From (\ref{eq:pout-2}), we invoke the identity $_{\phantom{}2}F_{1}\left(\alpha;\beta;\gamma;0\right)=1,$
(when $\eta\rightarrow\infty$) to obtain an asymptotic approximation
as
\begin{equation}
P_{\textrm{out}}^{asy}\backsimeq\frac{\Gamma\left(Nm+Nm_{s}\right)}{\Gamma\left(1+Nm\right)\Gamma\left(Nm_{s}\right)}\left(\frac{\gamma_{\textrm{th}}m}{\eta Nm_{s}}\right)^{Nm},\label{eq:pout-asympt}
\end{equation}
where it can be readily observed that the diversity gain is equivalent
to $Nm.$

It is worth noting that the asymptotic expressions for the capacity,
BER and outage probability in (\ref{eq:c-av-asympt-2}), (\ref{eq:av-ber-asym-2})
and (\ref{eq:pout-asympt}) respectively, have more tractable closed-forms
that are easier for analysis. As will be demonstrated in the next
section, these expressions closely approximate the exact solutions. 

\section{\textcolor{black}{Numerical Results and Discussions\label{sec:Results}}}

\textcolor{black}{In this section, we present and discuss results
from the mathematical expressions derived in the paper. We then investigate
the effect of key parameters on the system. The results are then verified
using Monte Carlo simulations with at least $10^{5}$ iterations.
Unless otherwise stated, }we have assumed RIS-to-$D$ distance has
been normalised to unit distance, $r_{D}=1$ m, pathloss exponent
$\beta=2.7$ and shadowing parameter $m_{s}=5$ (moderate).

\begin{figure}[th]
\begin{centering}
\textcolor{black}{\includegraphics[scale=0.45]{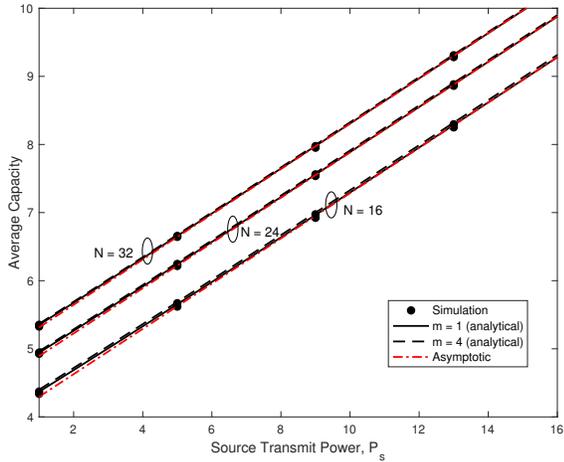}}
\par\end{centering}
\textcolor{black}{\caption{Average capacity versus source transmit power $P_{s}$ for the RIS-enabled
access point node. Parameters considered with varying severity of
fading parameter $m$ and number of RIS cells $N$.\label{fig:cs-vs-ps}}
}
\end{figure}

In Fig. \ref{fig:cs-vs-ps}, we commence analysis for the IoT RIS-enabled
network, with the average capacity against source power for different
numbers of RIS cells and fading parameters. We consider fading parameters
$m=1$ (heavy) and $m=4$ (moderate). It can be observed that the
average capacity at a fixed SNR value, is proportional to the number
of RIS cells. It can be further noted that within the region considered,
the fading parameter has a far lesser effect on the average capacity
compared to the source power and the number of RIS cells. Moreover,
the asymptotic solution closely matches the exact solutions. An observation
on doubling the number of cells from $N=16$ to 32 cells (at fixed
transmission power) corresponds to about 1 bit/Hz increase in average
capacity. Thus, from a design perspective, this may be insightful
for a cost-benefit decision.\textcolor{black}{}
\textcolor{black}{}
\begin{figure}[th]
\begin{centering}
\textcolor{black}{\includegraphics[scale=0.45]{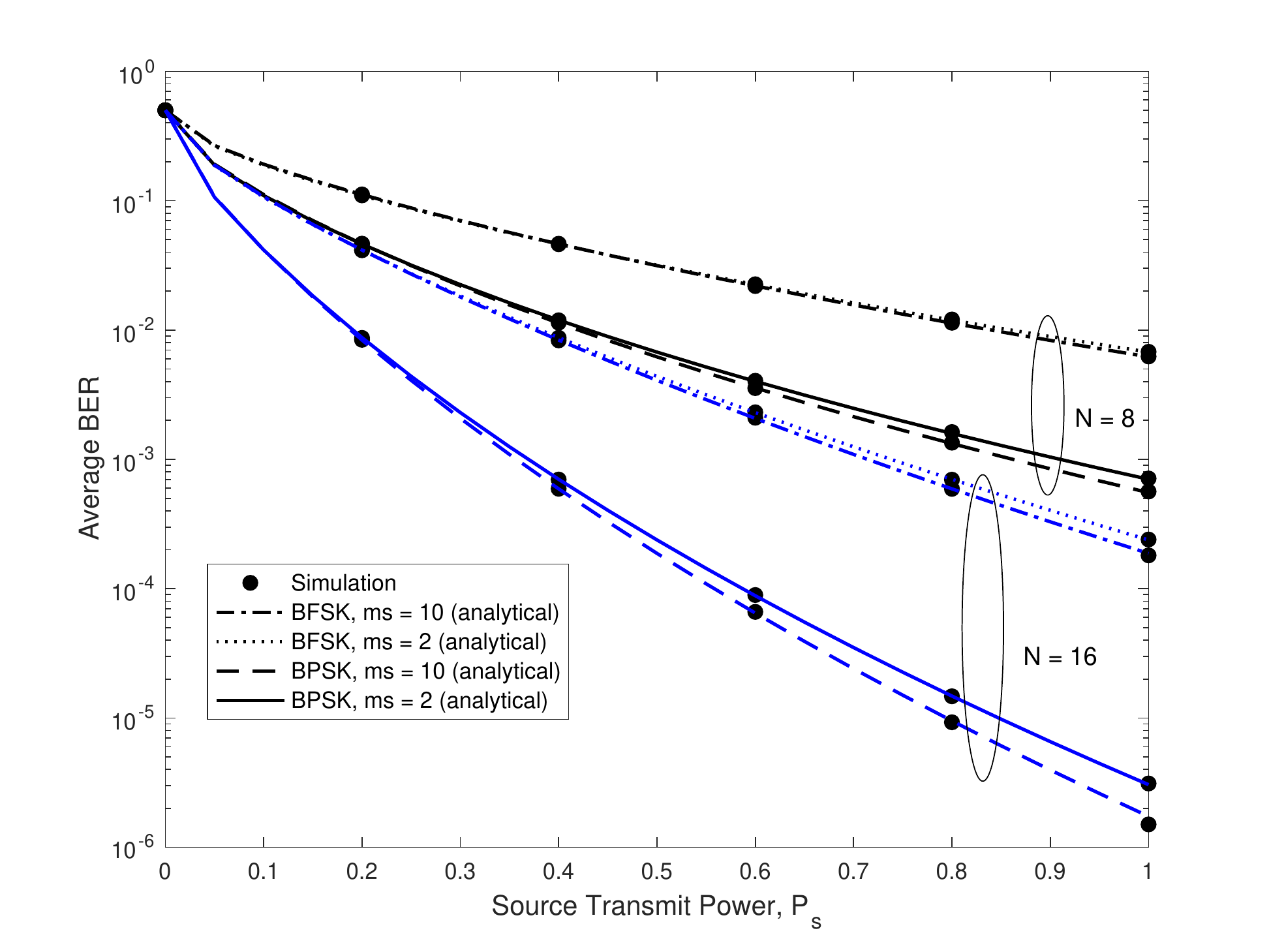}}
\par\end{centering}
\textcolor{black}{\caption{Average BER versus source transmit power $P_{s}$ for the RIS-enabled
access point node. Parameters considered with varying severity of
shadowing parameter $m_{s}$, number of RIS cells $N$ and modulation
scheme.\label{fig:ber-vs-ps}}
}
\end{figure}

Fig. \ref{fig:ber-vs-ps} shows a plot of the average BER against
source power for different numbers of RIS cells and fading parameters
at very low SNR (maximum 0 dB). We consider modulating schemes $\lambda=0.5$
(BFSK) and $\lambda=1$ (BPSK). We assume fading parameters $m=1$.
Overall, we observe that the average BER for both modulation schemes
is lower with a higher number of RIS cells $N=16$, except at very
low SNR, where BFSK $N=16,$ almost equal to best performing schemes
for $N=8$ for BPSK. Additionally, the improvement in the error performance
between the modulation schemes (and shadowing severities) is more
noticeable with higher number of cells at $N=16.$ The results also
demonstrates that the BPSK scheme performs better than BFSK for the
different observed regions of shadowing severities, source transmit
powers and number of RIS cells considered. We also note the clear
benefit of employing the RIS schemes by observing that, by doubling
the cells from $N=8$ to 16, we can achieve as an example, a BER improvement
of about 3 orders of magnitude for BPSK $m_{s}=2.$ The knowledge
of the effect of shadowing is particularly important in IoT networks,
since it is possible for the shadowing severity on a device to rapidly
change especially indoors.\textcolor{black}{}
\begin{figure}[th]
\begin{centering}
\textcolor{black}{\includegraphics[scale=0.45]{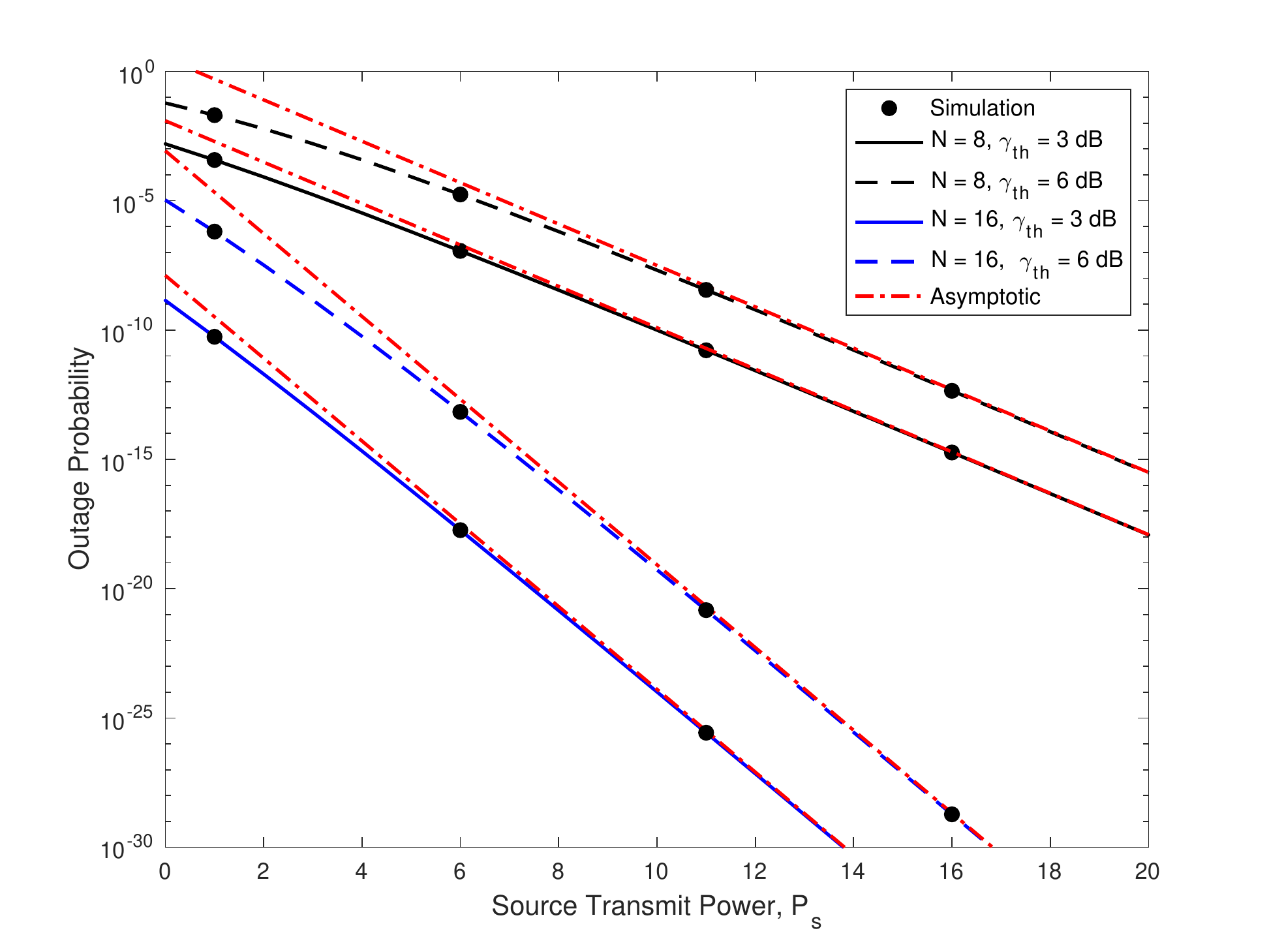}}
\par\end{centering}
\textcolor{black}{\caption{Outage probability versus the source transmit power $P_{s}$ for the
RIS-enabled access point node. Parameters considered for different
SNR threshold $\gamma_{\textrm{th}}$ and number of RIS cells $N$.
Fixed values $m=1$ and $m_{s}=5.$ \label{fig:OP-vs-threshold}}
}
\end{figure}

In Fig. \ref{fig:OP-vs-threshold}, we present the outage probability
against source transmit power for different number of RIS cells and
SNR thresholds. We assume a fading severity $m=1$ and shadowing parameter
$m_{s}=5$. It can be observed that for the number of RIS cells considered
at different transmit powers, acceptable levels of outage probabilities
can be achieved, even at low SNR regimes. It can be further observed
that the benefit of the RIS configuration indicates that as low as
0 dB, an improvement of at least 5 orders of magnitude can be achieved
by doubling the RIS cells from $N=8$ to 16, with further improvements
for higher SNR levels. Meanwhile, increasing the desired SNR threshold
by 3 orders of magnitude from $\gamma_{\textrm{th}}=3$ dB to 6 dB,
corresponds to an approximate increase of 2 to 4 orders of magnitude
in outage probability, which is in line with the diversity gain of
$N*m$ as earlier discussed. Thus, considering the cost involved,
it may be worth considering a transmit power increase, while adjusting
the desired SNR threshold for an approximately similar benefit. Furthermore,
plots of the asymptotic outage probability (obtained from (\ref{eq:pout-asympt})),
demonstrate that for both RIS-conifigurations considered in Fig. \ref{fig:OP-vs-threshold},
an SNR level of 9 dB to 10 dB is sufficient to allow convergence between
the asymptotic approximation and exact analytical expression in (\ref{eq:pout-2}). 

\section{\textcolor{black}{Conclusions\label{sec:Conclusions}}}

\textcolor{black}{In this paper, we examined }an IoT network employing
a reconfigurable intelligent surface (RIS) over composite fading and
shadowing channels. Exact expressions were derived for the efficient
computation of system metrics such as the average capacity, average
BER and outage probability, as well as approximate more tractable
asymptotic expressions. The effects of parameters such as source transmit
power, severity of fading and shadowing as well as number of RIS-cells
were investigated. The results indicate the clear benefit of employing
a RIS-enabled access point, with regards to the average capacity,
BER and outage probability, as well as the versatility of the derived
expressions in analysing the effects of fading and shadowing on the
network. The results further demonstrate that for a RIS-enabled IoT
network, there is the need to balance between the cost and benefit
of increasing the RIS cells against other parameters such as increasing
transmit power, especially at low SNR and/or high to moderate fading/shadowing
severity.

\bibliographystyle{IEEEtran}
\bibliography{bibGC19}

\end{document}